\documentclass[aps,pra,twocolumn,showpacs,superscriptaddress]{revtex4-1}
\usepackage{amsmath,amssymb}
\usepackage[matha,mathx]{mathabx}
\usepackage{txfonts}
\usepackage{float}
\usepackage[pdftex]{graphicx}
\usepackage{subfigure}
\usepackage{natbib}
\usepackage{bm}
\usepackage{braket}
\newcommand \beq{\begin{eqnarray}}
\newcommand \eeq{\end{eqnarray}}

\begin{document}

\title{Unconventional superfluidity in quasi-one dimensional systems}

\author{Shun Uchino}
\affiliation{DPMC-MaNEP, University of Geneva, 24 Quai Ernest-Ansermet,
CH-1211 Geneva, Switzerland}
\author{Akiyuki Tokuno}
\affiliation{DPMC-MaNEP, University of Geneva, 24 Quai Ernest-Ansermet,
CH-1211 Geneva, Switzerland}
\author{Thierry Giamarchi}
\affiliation{DPMC-MaNEP, University of Geneva, 24 Quai Ernest-Ansermet,
CH-1211 Geneva, Switzerland}

\date{\today}

\begin{abstract}
 We show that an unconventional superfluid triggered by spin-orbit coupling
 is realized for repulsively interacting quasi-one-dimensional fermions.
 A competition between spin-singlet and -triplet pairings
 occurs due to the breaking of inversion symmetry. We show that
 both superfluid correlations decay algebraically with the same exponent
 except for special coupling constants for which a dominant superfluid is
 controlled  by the spin-orbit coupling.
 We also discuss a possible experiment to observe such phases with
 cold atoms.
\end{abstract}


\maketitle

\section{Introduction}
Fermionic superfluids (SFs) ---superconductors for charged particles---
are ubiquitous in diverse fields ranging from condensed matter to
elementary particle physics. The first realizations of such systems
are  described by the $s$-wave spin-singlet pairing picture of the
celebrated BCS theory.
Alternatively, fermionic SFs, which cannot be described
by the standard BCS scenario, are called unconventional SFs.
Such systems are quite rare and since the
example of SF $^3$He ~\cite{RevModPhys.47.331} there have been
very few new candidates.
Unconventional superfluidity or superconductivity
is thus one of the extremely challenging topics in many-body physics.

This interest has been revived with the arrival
of cuprates as high-$T_c$ superconductors~\cite{anderson1997theory}.
Superconductivity in non-centrosymmetric systems ~\cite{bauer2012non}
has provided a novel route to unconventional superconductors. In these systems
an admixture of spin-singlet and -triplet pairings is believed to be realized
due to spin-orbit coupling (SOC) attributed to the breaking of inversion
symmetry~\cite{SovPhysJETP.68.1244.1989,PhysRevLett.75.2004,PhysRevLett.87.037004,PhysRevLett.92.097001}.
However the presence of strong correlations in this class of materials
makes it difficult to analyze theoretically such an admixture with a
well defined starting point.

When dealing with strong correlations, one or quasi-one dimensional
systems provide some insight into their counterparts in higher
dimensions by allowing the use of powerful analytical and numerical
approaches~\cite{giamarchi2003quantum}.
For a standard SF, it is exactly known that
the evolution from the BCS to Bose-Einstein condensation (BEC)
is continuous by using Bethe
ansatz~\cite{PhysRevLett.93.090405,PhysRevLett.93.090408}.
For unconventional SFs, the suggestion
of a resonant valence bond mechanisms~\cite{anderson1997theory} and
non-Fermi liquids~\cite{jiang2012non} has been investigated for fermionic
ladders. It has been established that in a ladder, a spin-singlet SF is
present with purely repulsive interactions~\cite{PhysRevB.47.10461,PhysRevB.48.15838,PhysRevB.50.252,nagaosa1995bipolaron,
PhysRevB.53.R2959,PhysRevB.53.12133}.

In addition to the realizations in condensed matter,
ultracold atomic gases have offered an ideal playground for studying
strongly interacting SFs due to the high controllability
of the microscopic parameters and Hamiltonians~\cite{RevModPhys.80.885}.
By tuning an interatomic interaction with a Feshbach resonance,
the BCS-BEC crossover has been achieved for two component
fermions~\cite{zwerger2011bcs}.
To go beyond the standard BCS scenario, experimental
efforts~\cite{zwierlein2006fermionic,partridge2006pairing,PhysRevLett.97.190407,shin2008phase,PhysRevLett.103.170402,liao2010spin}
are devoted to realizing $p$-wave SFs by using a $p$-wave Feshbach
resonance~\cite{RevModPhys.82.1225} and
Fulde-Ferrell-Larkin-Ovchinnikov~\cite{PhysRev.135.A550,larkin1965inhomogeneous}
SFs by preparing spin-imbalanced gases.
In addition, currently available synthetic gauge fields~\cite{PhysRevLett.102.046402,lin2009synthetic,lin2011synthetic,lin2011spin,PhysRevLett.107.255301,PhysRevLett.108.225303,PhysRevLett.108.225304,PhysRevLett.109.095301,PhysRevLett.109.095302,PhysRevLett.109.115301,2013arXiv1304.5520S,RevModPhys.83.1523,aidelsburger2013experimental},
where electromagnetic fields and SOC can be mimicked
in both continuum and lattice spaces
by using Raman lasers and driving an optical lattice,
pave the way for the realization of nontrivial
SFs in cold atoms. In fact, some theoretical works point out that
such SFs emerge in higher
dimensions~\cite{PhysRevB.83.094515,PhysRevB.84.014512,PhysRevLett.107.195303,PhysRevLett.107.195304,PhysRevLett.107.195305,PhysRevA.84.063618}.
However, it is also a challenging issue to demonstrate SFs with
an admixture between spin singlet and triplet pairings
in strongly-correlated optical lattice systems.

\begin{figure}[t]
 \begin{center}
 \includegraphics[width=7.5cm]{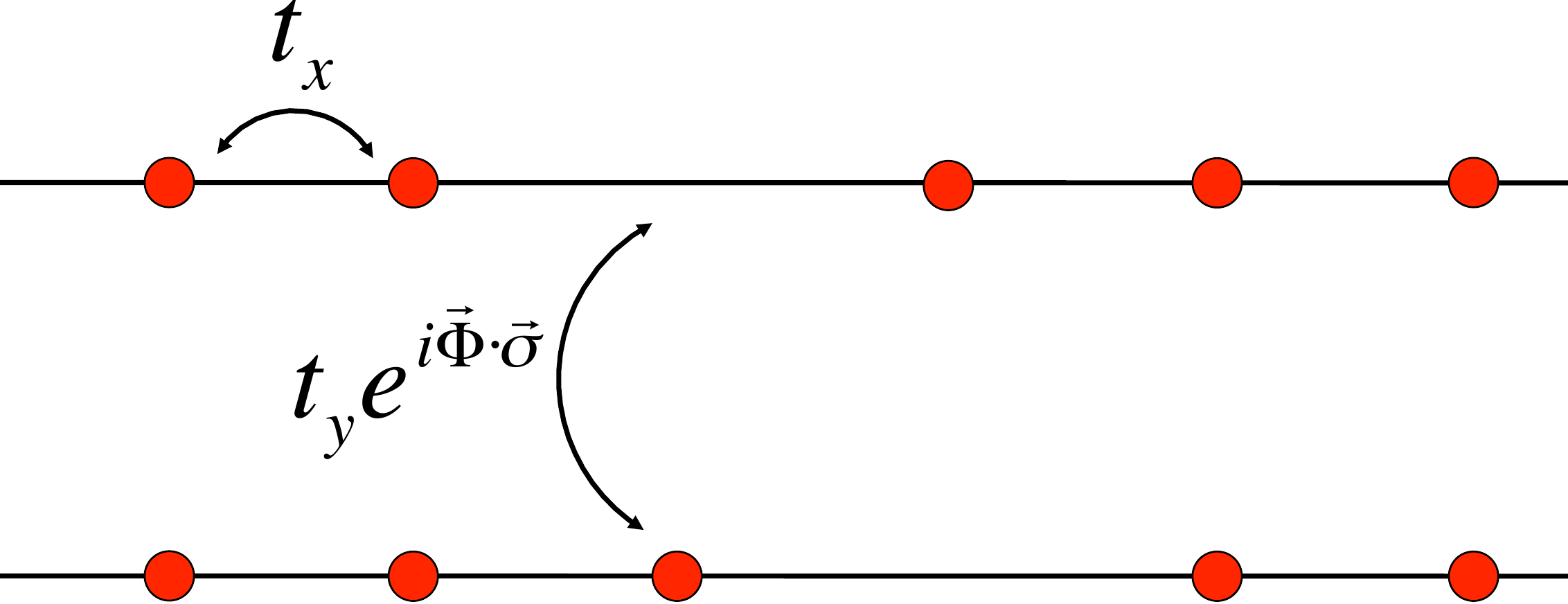}
  \caption{(Color online)
  Two-coupled fermionic ladder system with a SOC along the rung
  direction.
  Due to the SOC, the rung hopping term takes a complex value and
  depends on the spin.}
 \label{fig:hamiltonian}
 \end{center}
\end{figure}

In this paper, we propose a novel system of two fermionic chains
with  SOCs, which breaks inversion symmetry.
As a result, an unconventional SF with an admixture between spin singlet
and triplet pairings emerges,
which originates from the repulsive interaction and whose
mechanism thus differs from a standard BCS scenario for
attractive atom gases with SOCs in continuum space.
Since such an admixture is
controlled by the SOC,
this model can also be used to realize an ideal spin-triplet SFs in cold atoms.
Furthermore, we show by using a mapping between attractive and repulsive
situations that our findings are directly relevant for present experiments
and discuss the observable consequences.

\section{Model}
We consider two coupled fermionic chains with SOCs.
There are two different way to include the SOC effect: along
the chain and along rung directions.
The SOC along the chain  direction corresponds to a
modification of the boundary condition of the
chains~\cite{PhysRevLett.7.46} as long as the effect of the SOC
is equal in the two chains~\footnote{If the effect of the SOC is not
equal in the two chains, non-trivial states may emerge since such a
effect introduces a spin flux in the system. As for a charge flux, see,
e.g., G. Roux, E. Orignac, S. R. White, and D Poilblanc, Phys. Rev. B
\textbf{76}, 195105 (2007);
A. Jaefari and E. Fradkin, Phys. Rev. B \textbf{85}, 035104 (2012).}.
Thus we focus on the case where the SOC is applied to the rung
direction (Fig.~\ref{fig:hamiltonian}), and the Hamiltonian is given as
\beq
H^{(c)}_{\Phi}
=-t_{x}\sum_{j,p,\sigma}
(c^{\dagger}_{j,p\sigma}c_{j+1,p\sigma}+h.c.)
+U\sum_{j,p}n_{j,p\uparrow}n_{j,p\downarrow}
\nonumber\\
-t_{y}\sum_{j,\sigma,\sigma'}
((e^{i\vec{\Phi}\cdot\vec{\sigma}})_{\sigma,\sigma'}
c^{\dagger}_{j,1\sigma}c_{j,-1\sigma'}
+h.c.),
\label{eq:hamiltonian}
\eeq
where $p=\pm 1$ and $j$ are chain and site indices, respectively.
The vector $\vec{\sigma}=(\hat{\sigma}_x,\hat{\sigma}_y,\hat{\sigma}_z)$
denotes Pauli matrices.
The Peierls phase $\vec{\Phi}\cdot\vec{\sigma}$ acts on the spin sector,
and the direction of $\vec{\Phi}$ determines the type of the SOC:
For $\vec{\Phi}$ parallel and perpendicular to a spin-quantization axis,
the SOC effect works as a spin-dependent hopping and as a spin-flip
hopping, respectively.
We note that in the presence of the SOC,
inversion symmetry along the rung direction is explicitly broken.
The parameters $t_x$, $t_y$ and $U$, respectively, denote intra- and
inter-chain hopping, and the Hubbard interaction.
In this paper we restrict ourselves to be in the following parameter
regime in which the two-band nature in the ladder
is relevant in the low-energy
physics:
\beq
t_x\gg t_y,\,U \quad \mbox{or} \quad t_x\approx t_y\le U.
\label{eq:condition}
\eeq
In addition, we consider only the incommensurate case for which a gapless
charge excitation exists.

We briefly review the physics of a repulsively interacting fermionic
ladder for $\vec{\Phi}=\vec{0}$.
In the presence of the rung hopping $t_{y}$, the two energy bands coming
from the chains are split, and the two-band structure and their
competition become crucial for the low-energy physics as long as the
system is in the parameter regime~(\ref{eq:condition}).
Correlation of inter-chain spin-singlet SF are
dominant.~\footnote{For the general phase diagram in the fermionic
ladder system, see, e.g. \cite{giamarchi2003quantum}, and references therein.}
This result is well-established from the
analytical~\cite{PhysRevB.47.10461,PhysRevB.48.15838,PhysRevB.50.252,nagaosa1995bipolaron,PhysRevB.53.R2959,PhysRevB.53.12133,giamarchi2003quantum}
and numerical
viewpoint~\cite{PhysRevB.53.11721,noack1996ground,PhysRevB.65.165122,PhysRevB.75.245119}.
The inter-chain SF operator relevant for the repulsion is given by
\beq
O^{(c)}_{\text{SSC}}=c_{j,1\uparrow}c_{j,-1\downarrow}
-c_{j,1\downarrow}c_{j,-1\uparrow},
\label{eq:SSC_operator}
\eeq
which is also called $d$-wave-like SF operator because the symmetry
resembles the $d_{x^2-y^2}$ pairing~\cite{PhysRevB.53.R2959}
where in Fourier space, 
the components with transverse wave vector, 0 and $\pi$ 
have opposite signs, 
and thus the sign of the operator changes with a $\pi/2$ rotation..

\section{Unconventional superfluid in the presence of spin-orbit coupling}
We now show the emergence in this model of an unconventional SF induced
by the SOC. As mentioned, we have different types of SOC depending on the
vector $\vec{\Phi}$, and we discuss the three orthogonal cases:
$\vec{\Phi}_z=(0,0,\Phi)$,
$\vec{\Phi}_y=(0,\Phi,0)$, and $\vec{\Phi}_x=(\Phi,0,0)$ where
$-\pi<\Phi\le\pi$.

Let us start with $\vec{\Phi}=\vec{\Phi}_z$.
We use the following canonical transformation
\beq
c_{j,1\sigma}= d_{j,1\sigma},
\ c_{j,-1\sigma} =
e^{-i\sigma\Phi}d_{j,-1\sigma},
\label{eq:trans1}
\eeq
which twists the spins in the $p=-1$ chain around
the $\vec{\Phi}_z$ direction.
Due to Eq.~(\ref{eq:trans1}), the phase $e^{-i\sigma\Phi}$
is absorbed into $d_{j,p\sigma}$, and thus the form of
the Hamiltonian~(\ref{eq:hamiltonian}) becomes $H^{(d)}_{\Phi=0}$ identical, 
in term of the operators $d_{j,p\sigma}$ to the original one~(\ref{eq:hamiltonian}) 
but for $\Phi=0$. Thus the present problem is mapped onto one without the
SOC, for which one can directly use the above solution for fermionic ladders. 
The ground-state in the $d$ representation in our situation,
incommensurate filling and parameter regime~(\ref{eq:condition}), is
thus dominated by the inter-chain SF pair correlation represented by the
operator~(\ref{eq:SSC_operator}).

\begin{figure}[t]
 \begin{center}
 \includegraphics[width=7.5cm]{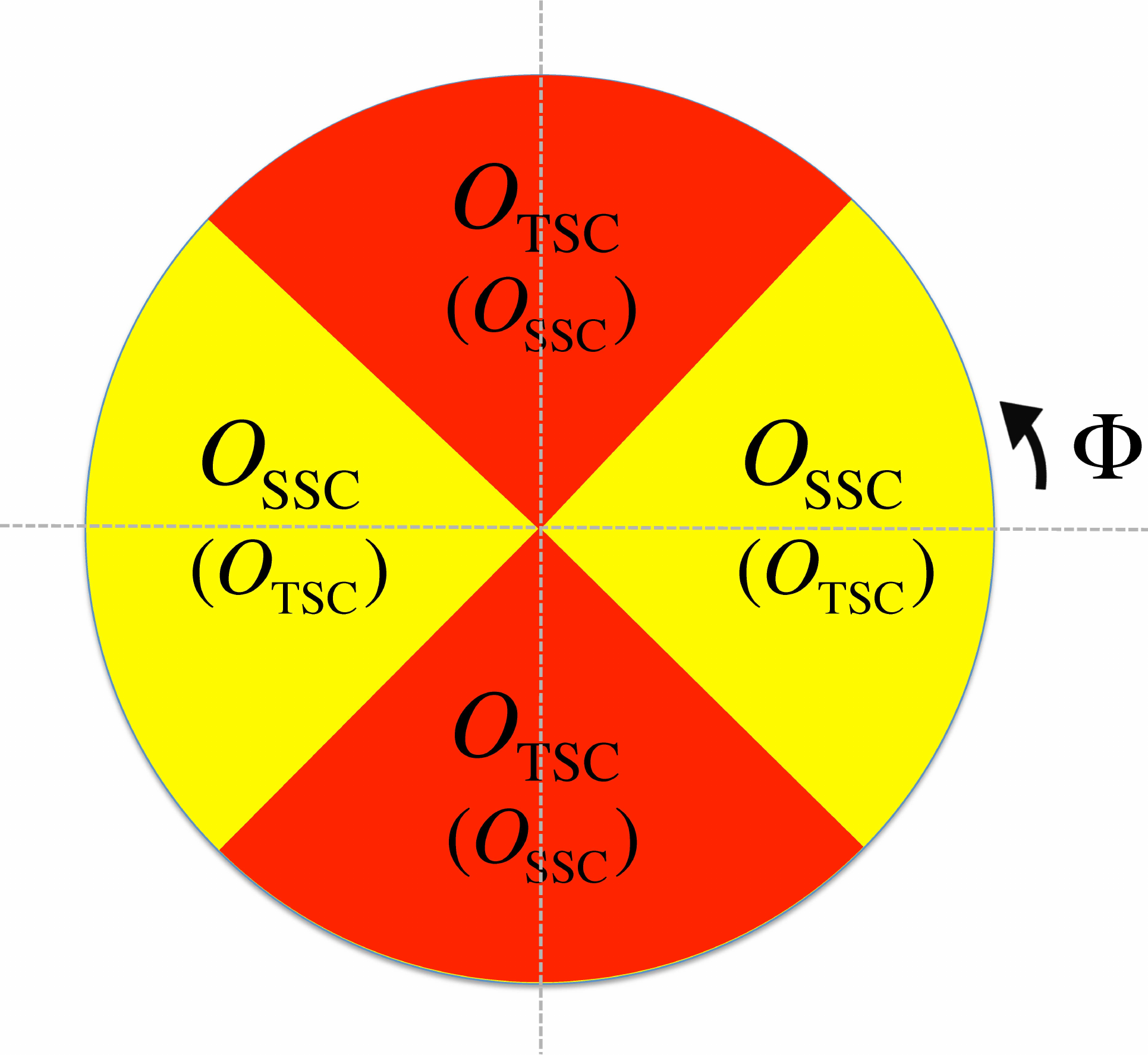}
 \caption{(Color online)
  Phase diagram as a function of the SOC parameter $\Phi$ for a repulsive
  Hubbard interaction $U$.
  The parentheses denotes subdominant fluctuations.
  The subdominant fluctuations disappear at $\Phi=0, \pm\pi/2, \pi$
  (dotted line), and
  both the pairing types are equally balanced at
  $\Phi=\pm \pi/4, \pm 3\pi/4$.}
 \label{fig:phase}
 \end{center}
\end{figure}

We next consider how the ground state in the $d$ representation is described in
the original $c$ representation via the transformation~(\ref{eq:trans1}).
In addition to the spin-singlet SF~(\ref{eq:SSC_operator}) we
also look at the inter-chain spin-triplet SF along the $z$
direction, which is represented as
\beq
O^{(c)}_{\text{TSC}^z}=c_{j,1\uparrow}c_{j,-1\downarrow}
+c_{j,1\downarrow}c_{j,-1\uparrow}.
\eeq
Due to Eq.~(\ref{eq:trans1}), the operators
$O^{(c)}_{\text{SSC}}$ and $O^{(c)}_{\text{TSC}^z}$ are transformed as
follows:
\beq
O^{(c)}_{\text{SSC}}=\cos\Phi O^{(d)}_{\text{SSC}}+i\sin\Phi
O^{(d)}_{\text{TSC}^z},\\
O^{(c)}_{\text{TSC}^z}=\cos\Phi O^{(d)}_{\text{TSC}^z}+i\sin\Phi
O^{(d)}_{\text{SSC}}.
\eeq
Note that both the spin-singlet and -triplet SF operators share
$O^{(d)}_{\text{SSC}}$, and there is no other operator, including
$O^{(d)}_{\text{SSC}}$, in the $c$ representation.
Therefore, recalling that the $O^{(d)}_{\text{SSC}}$ correlation is
dominant in the $d$ representation, the asymptotic form of the correlation of
$O^{(c)}_{\text{SSC}}$ and $O^{(c)}_{\text{TSC}^z}$ is written as
\beq
\langle O^{(c)\dagger}_{\text{SSC}}(r) O^{(c)}_{\text{SSC}}(0)\rangle_c
\sim \cos^2\Phi
\langle O^{(d)\dagger}_{\text{SSC}}(r) O^{(d)}_{\text{SSC}}(0)\rangle_d,
\label{eq:ssc}\\
\langle O^{(c)\dagger}_{\text{TSC}^z}(r) O^{(c)}_{\text{TSC}^z}(0)\rangle_c
\sim \sin^2\Phi
\langle O^{(d)\dagger}_{\text{SSC}}(r) O^{(d)}_{\text{SSC}}(0)\rangle_d,
\label{eq:tsc}
\eeq
where $\langle \cdots \rangle_{c(d)}$ represents correlations
in the $c$ ($d$) representations.
As can be seen from Eqs.~\eqref{eq:ssc} and~\eqref{eq:tsc},
the two different correlations in the original
representation are described by the same asymptotic form except for the
prefactor given by the SOC parameter $\Phi$.
Since the algebraic decaying function
$\langle O^{(d)\dagger}_{\text{SSC}}(r)O^{(d)}_{\text{SSC}}(0)\rangle_d$
is independent of $\Phi$, the relevancy of
$\langle O^{(c)\dagger}_{\text{SSC}}(r) O^{(c)}_{\text{SSC}}(0)\rangle_c$
and
$\langle O^{(c)\dagger}_{\text{TSC}^z}(r) O^{(c)}_{\text{TSC}^z}(0)\rangle_c$
is determined solely by the coefficients $\cos^2\Phi$ and $\sin^2\Phi$.
For example, the inter-chain spin-singlet and -triplet pairing
are dominant, respectively, for $0\le\Phi<\pi/4$ and for
$\pi/4<\Phi\le\pi/2$.
The ideal spin-singlet and -triplet pair states are realized only at
$\Phi=0$ and $\Phi=\pi/2$, respectively, and the spin-singlet and
-triplet pairings are equally mixed at $\Phi=\pi/4$.
In Fig.~\ref{fig:phase} we show the phase diagram as a function of $\Phi$.

In the case of $\vec{\Phi}=\vec{\Phi}_{y}$,
we first implement a global spin rotation as
\beq
f_{j,p\sigma}
=\sum_{\eta,\eta'}
 \left[e^{-i\frac{\pi}{4}\hat{\sigma}_{x}}\right]_{\eta,\eta'}
 c_{j,p\eta'}.
\label{eq:trans2}
\eeq
The Hamiltonian is then transformed into the one for
$\vec{\Phi}_z$, which was already studied above.
By using the same transformation (\ref{eq:trans1}) for
$f_{j,p\sigma}$, the Hamiltonian is reduced again to the one
without the SOC. The fermion operators in the consequent
Hamiltonian is again  $d_{j,p,\sigma}$.
As a result the following fluctuation operators in the original $c$
representation are found to include the most relevant $O^{(d)}_{\mathrm{SSC}}$ in the $d$
representation:
\beq
O^{(c)}_{\text{SSC}}=\cos\Phi O^{(d)}_{\text{SSC}}+i\sin\Phi
O^{(d)}_{\text{TSC}^z},\\
O^{(c)}_{\text{TSC}^{y}}= i\cos\Phi O^{(d)}_{\text{TSC}^z}-\sin\Phi
O^{(d)}_{\text{SSC}},
\eeq
where the spin-triplet SF operator along the $y$  direction
has been defined as
\beq
O^{(c)}_{\text{TSC}^{y}}=c_{j,1\uparrow}c_{j,-1\uparrow}
+ c_{j,1\downarrow}c_{j,-1\downarrow}.
\eeq
As in the case of $\vec{\Phi}_z$, the dominant fluctuations are found to
be the spin-singlet and -triplet SF pair as a function of $\Phi$, and their weight
are determined by $\Phi$, as illustrated in Fig.~\ref{fig:phase}.
A similar analysis can be applied for
$\vec{\Phi}=\vec{\Phi_x}$, and find that
the spin-triplet SF along the $x$ direction competes with
the spin-singlet SF.

We can also discuss an arbitrary $\vec{\Phi}$ case using the above
result. 
Taking into account the SU(2) spin rotational symmetry in the
Hamiltonian of each chain, the appropriate spin rotation reduces the
general $\vec{\Phi}$ problem to the $\vec{\Phi}_z$ one, as seen in the
discussion in the $\vec{\Phi}_{x}$ and $\vec{\Phi}_{y}$ cases.
Thus, admixture of the spin singlet and triplet pairings is concluded to be
also obtained in an arbitrary direction of the SOC. 

To summarize, unconventional spin-triplet SF pairs are found to
be induced by the SOC, accompanying spin-singlet SF
pairs, which occurs for the {\it repulsive interaction}.
The weight of the spin-triplet pair is determined by the magnitude of
the SOC parameters. The direction of
the spin-triplet pairs, which is  the so-called $d$ vector
in a spin-triplet $p$-wave SF \cite{RevModPhys.47.331},
corresponds to the one of the SOC vector $\vec{\Phi}$.
The results are summarized in Table~\ref{table}.
The most important point is that one can control the
unconventional SF pair through the SOC.

Let us discuss the physical reason of the emergence of spin-triplet SF pairs.
As mentioned above, the canonical transformation~(\ref{eq:trans1})
physically rotates spins on the $p=-1$ chain around
$\vec{\Phi}_z$ by $\Phi$, which is a direct way to change spin
singlets to spin triplets.
This can be easily confirmed by operating the spin rotation to one spin
of a pair, $Z_2(\theta)=e^{i\theta\hat{\sigma}_{z}}$,
to a singlet spin pair
$\ket{s}=\frac{1}{\sqrt{2}}[\ket{\uparrow}_1\ket{\downarrow}_2-\ket{\downarrow}_1\ket{\uparrow}_2]$:
For $\theta=\pi/2$, it gives
$Z_2(\theta)\ket{s}=-\frac{i}{\sqrt{2}}[\ket{\uparrow}_1\ket{\downarrow}_2+\ket{\downarrow}_1\ket{\uparrow}_2]$
which is exactly a triplet spin pair.
In other words the SOC we considered 
rotates one spin of the inter-chain spin pair.
Therefore, one concludes that the essential mechanism of the
SF is identical to the inter-chain spin-singlet SF in the
conventional fermionic ladder, and the one spin twisting by the SOC
transforms the spin-singlet pair into a spin-triplet
one \footnote{
While we can consider the same model with attractive $U$ at incommensurate
filling, where on-site spin-singlet SF correlation is dominant
without the SOC, this on-site SF is always the dominant fluctuation
in the presence of the SOC. Namely, admixture between spin singlet and
triplet pairings does not occur in this case.}.

\begin{table}
\caption{Relationship between the direction of the SOC and
competing interchain spin triplet and singlet pairings.}
\label{table}
\begin{center}
\begin{tabular}{l|c|c}
\hline
Direction & Spin-triplet pairing & Spin-singlet pairing  \\
\hline
$\vec{\Phi}_x=(\Phi,0,0)$ & $O^{(c)}_{\text{TSC}^x}$  &   \\
$\vec{\Phi}_y=(0,\Phi,0)$ & $O^{(c)}_{\text{TSC}^y}$ & $O^{(c)}_{\text{SSC}}$ \\
$\vec{\Phi}_z=(0,0,\Phi)$ & $O^{(c)}_{\text{TSC}^z}$ &  \\
\hline
\end{tabular}
\end{center}
\end{table}

Finally let us comment on effects of disorder, on the unusual SF
described above. For the system without the SOC,
an analysis with bosonization and
renormalization group methods predicts that
in the presence of non-magnetic disorder,
the inter-chain SF realized for $U>0$ is vulnerable
to disorder while the intra-chain SF realized for $U<0$ is
much more stable~\cite{PhysRevB.56.7167}.
Since a non-magnetic disorder favors SU(2) symmetry,
these properties survive in the presence of the SOC.
This is interpreted as the one dimensional analog of the fact
that an isotropic $s$-wave SF is robust against disorder
(Anderson's theorem)
while an anisotropic SF could be destroyed by disorder.

\section{attractive route }
\begin{table}
\caption{Correspondence among different quantities
under the particle-hole transformation.}
\label{table2}
\begin{center}
\begin{tabular}{c|c}
\hline
Original repulsive model & Corresponding attractive model \\
\hline
Away from half filling & Half filling \\
Spin balance &  Spin imbalance \\
SOC along the $z$ direction & U(1) gauge field in the charge sector  \\
\hline
\end{tabular}
\end{center}
\end{table}

We now discuss the experimental possibilities to probe
the unconventional SF predicted here.
Because of the difficulty of confirming SFs directly in
one-dimensional fermionic optical lattices,
we alternatively employ the idea of the so-called attractive route to
indirectly check it with a particle-hole
transformation~\cite{PhysRevA.79.033620}.
We first consider the $\vec{\Phi}=\vec{\Phi}_z$ case.
The particle-hole transformation is well-defined on a bipartite lattice
as
\beq
c_{j,p\uparrow}= h_{j,p\uparrow}, \
c_{j,p \downarrow}= (-1)^{j+\Theta(-p)}
h^{\dagger}_{j,p\downarrow},
\label{eq:ph-trans}
\eeq
where $\Theta$ is the Heaviside step function.
Due to Eq.~(\ref{eq:ph-trans}),
$n_{j,p\downarrow}\to -n_{j,p\downarrow}$,
and the inter-chain hopping changes as
\beq
-t_{y}\sum_{j,\sigma}
(e^{i\Phi}h^{\dagger}_{j,1\sigma}h_{j,-1\sigma}+h.c.).
\eeq
Therefore the transformation~(\ref{eq:ph-trans}) maps the
Hamiltonian~(\ref{eq:hamiltonian}) on the one of the attractive fermionic
ladder with a U(1) gauge field along the rung direction.
While this type of U(1) gauge field already exists for $^{87}$Rb Bose
atom experiments~\cite{PhysRevLett.108.225304,2013arXiv1304.5520S},
this experimental technique can be applied in principle regardless
of atom statistics. In addition, the
transformation~(\ref{eq:ph-trans}) exchanges the roles of the particle
filling and magnetization of a state:
spin-imbalanced states at half-filling is transformed into spin-balanced
away from half-filing, and vice versa. (See Table II.)
The attractive system after the mapping by
Eq.~(\ref{eq:ph-trans}) is thus potentially easily realizable in experiments,
since the filling, one particle per site, is more naturally formed around a
trap center by tuning trapping frequency and particle
number.

From Eq.~(\ref{eq:ph-trans}), the inter-chain spin-singlet
SF is changed to
\beq
\text{Re,Im}(O^{(c)}_{\text{SSC}})=
(-1)^{j+1}\sum_{p\sigma}p(\sigma^{x,y})_{\sigma,-\sigma}h^{\dagger}_{j,p\sigma}
h_{j,-p-\sigma},\nonumber\\
\label{eq:corr1}
\eeq
which is an operator for a staggered spin flux phase on a plaquette
in which there is no global spin flux but a local spin flux.
This local staggered spin flux~\eqref{eq:corr1} is measurable
because the technique to observe local phases on a four-square plaquette
has been experimentally demonstrated in
Refs.~\cite{PhysRevLett.107.255301,aidelsburger2013experimental},
and a scheme for the spatially-resolved measurement of the
current is also proposed ~\cite{2013arXiv1309.3890K}.
On the other hand, the inter-chain spin-triplet SF
$O_{\text{TSC}^z}$ is changed to
\beq
\text{Re,Im}(O^{(c)}_{\text{TSC}^z})=
(-1)^{j+1}\sum_{p\sigma}(\sigma^{x,y})_{\sigma,-\sigma}h^{\dagger}_{j,p\sigma}
h_{j,-p-\sigma},\nonumber\\
\label{eq:corr2}
\eeq
which is an operator for a bond antiferromagnetic density wave along
the rung direction.
Such a correlation~\eqref{eq:corr2} may be measured
in a similar way than the recently implemented measurement of the
nearest-neighbor spin
correlations~\cite{PhysRevLett.105.265303,greif2013short}.

Next we consider the case of $\vec{\Phi}=\vec{\Phi}_x$ or $\vec{\Phi}_y$.
Then it turns out that Eq.~(\ref{eq:ph-trans}) contains
$h_{j,1\uparrow}h_{j,-1\downarrow}$ from the rung hopping term,
in which the number of particles is not conserved.
Indeed, the off-diagonal spin operators, i.e., spin flip
hoppings, correspond to $\eta$ pairings which constitute the
off-diagonal ones of SU(2) algebra in the charge
sector~\cite{Essler.etal/book.2010}.
Since Hamiltonians in cold atoms conserve the total number of particles,
the attractive route would be much less useful in the
$\vec{\Phi}=\vec{\Phi}_x$ and $\vec{\Phi}_y$ cases.

Finally, we estimate the possible parameter regime to measure the
states predicted in this paper.
First we need a sufficiently low temperature, $T< t_x,t_y,U,\Delta$
where $\Delta$ means an energy scale of gaps characterizing the realized
state.
If we choose $t_x=t_y$ and $4<U/t_x\le 8$ as a specific case in the
parameter regime~(\ref{eq:condition}), this energy scale is
estimated to be of the order of the exchange energy
$\Delta=4t_x^2/U$~\cite{PhysRevB.53.11721,noack1996ground,PhysRevB.65.165122,PhysRevB.75.245119},
and thus the necessary temperature would be $\le 10^{-1}t_x$.
Let us note that this estimation works well even in
the presence of a trap potential.
From the point of view of the local density approximation,
the realized state can form around the trap center, which can be regarded
as bulk, while it should be broken on the edges. The contribution from 
such balk-like regime must thus be non-negligible for the state to be observed. 
This would be possible as long as the trapped system size is large enough, 
which should possible given the existing experiments~\cite{greif2013short}
and the proposal~\cite{PhysRevA.86.023606}.

\section{Summary}
We have analyzed two coupled fermionic chains with a repulsive
interactions and found that an unusual SF emerges because of SOCs along the
rung direction. The properties such as the dominance and the $d$
vector of a triplet pair are controllable.
We have also discussed how to observe experimentally such a SF state
via a particle-hole transformation and the use of attractive interactions.
A staggered spin flux and bond antiferromagnetic correlations are possibly measurable 
observables associated with our predicted SF state.
We finally point out that the discussion based on the canonical
transformation~(\ref{eq:trans1}) can be widely applied, e.g., to
systems with a long range interaction and with spin imbalance,
which allows for further explorations of unconventional quantum states.

\section*{acknowledgement}
This work was supported by the Swiss National Foundation under
MaNEP and division II.

\bibliographystyle{apsrev4-1}
%

\end{document}